\documentclass{iopart}
\usepackage{graphicx}
\begin{document}

\title{Influence of superohmic dissipation on a disordered quantum critical point}

\author{Thomas Vojta$^1$, Jos\'e A Hoyos$^2$, Priyanka Mohan$^3$ and Rajesh Narayanan$^3$.}

\address{$^1$ Department of Physics, Missouri University of Science and Technology, Rolla,
MO 65409, USA\\
$^2$ Instituto de F\'{\i}sica de S\~ao Carlos, Universidade de S\~ao Paulo,
C.P. 369, S\~ao Carlos, S\~ao Paulo 13560-970, Brazil\\
$^3$ Department of Physics, Indian Institute of Technology Madras, Chennai 600036, India.}
\eads{\mailto{vojtat@mst.edu},  \mailto{hoyos@ifsc.usp.br}, \mailto{priyanka@physics.iitm.ac.in}, \mailto{rnarayanan@physics.iitm.ac.in}.}

\begin{abstract}
We investigate the combined influence of quenched randomness and dissipation on a quantum critical point with
$O(N)$ order-parameter symmetry. Utilizing a strong-disorder renormalization group, we determine the critical behavior
in one space dimension exactly. For \emph{superohmic} dissipation, we find a Kosterlitz-Thouless type transition with conventional
(power-law) dynamical scaling. The dynamical critical exponent depends on the spectral density of the dissipative baths.
We also discuss the Griffiths singularities, and we determine observables.
\end{abstract}

\submitto{\JPCM}

\section{Introduction}
Real physical systems are seldom bereft of quenched disorder that occurs in the guise of impurities, defects or
other lattice imperfections. This disorder can have a significant impact on the physics near phase transitions.
These effects are particularly dramatic near zero-temperature quantum phase transitions (QPTs) where they give rise to
a bevy of fascinating phenomena such as quantum Griffiths effects \cite{ThillHuse95,RiegerYoung96},
infinite-randomness critical points featuring activated dynamical
scaling~\cite{Fisher92,Fisher95}, as well as smeared phase transitions~\cite{Vojta03a}
(for recent reviews, see, e.g., \cite{Vojta06,Vojta10}).

It is by now well understood that the enhanced disorder effects at quantum phase transitions can be attributed to
the disorder being perfectly correlated along the \emph{imaginary time} axis which becomes infinitely extended at
zero temperature. A particularly important role is played by the rare regions, droplets that are caught in the wrong
phase due to atypical disorder fluctuations (for instance, ferromagnetic droplets in a bulk paramagnet).
Due to the perfect disorder correlations in imaginary time, these rare regions are infinitely extended in the time
direction which severely hampers their dynamics and thus leads to the exotic phenomena mentioned above.

A dissipative environment further enhances these disorder effects. An example is provided by the magnetic quantum
phase transition in the random transverse-field Ising chain. In the absence of dissipation, this system features
an infinite-randomness quantum critical point separating the ferromagnetic and paramagnetic ground state phases.
It displays unconventional activated dynamical scaling and is accompanied by power-law quantum Griffiths singularities \cite{Fisher92,Fisher95}.
Ohmic dissipation prevents sufficiently large ferromagnetic droplets from tunneling
\cite{CastroNetoJones00,MillisMorrSchmalian01,MillisMorrSchmalian02}. This destroys the sharp quantum phase transition
by smearing, as was first predicted by phenomenological arguments \cite{Vojta03a} and later confirmed by
strong-disorder renormalization group calculations \cite{SchehrRieger06,SchehrRieger08,HoyosVojta08}.

Recently, some of us \cite{HoyosKotabageVojta07,VojtaKotabageHoyos09} considered the influence of Ohmic dissipation on
the quantum phase transition of a disordered rotor model with continuous $O(N)$ order parameter symmetry. Such a model appears
as the order parameter field theory of the itinerant antiferromagnetic transition \cite{Hertz76} or the superconductor-metal
transition in nanowires \cite{SachdevWernerTroyer04}.
By applying a strong-disorder renormalization group, it was shown that the critical behavior of this transition
is controlled by an infinite-randomness critical point that falls into the same universality class as the
random transverse-field Ising model. While this result may appear surprising at the first glance (because the two
systems have different symmetries), it follows from the fact that the rare regions in both systems are
exactly at the lower critical dimension of the problem, in agreement with the general classification scheme put
forward in Ref.\ \cite{VojtaSchmalian05}.

The case of Ohmic dissipation and the resulting exotic infinite-randomness critical point
should be contrasted with the dissipationless case, relevant, e.g., for dirty bosons \cite{AKPR04}, which features a
Kosterlitz-Thouless phase transition with conventional (power-law) dynamical scaling.
The qualitative difference between these two phase-transition scenarios immediately poses the question:
What happens in the presence of dissipation that is weaker than Ohmic and thus ``interpolates'' between
the two cases mentioned above?

In this paper we therefore study the quantum phase transition of a disordered $O(N)$ rotor model
in the presence of \emph{super-Ohmic} dissipation. We apply a strong-disorder renormalization group
to the order-parameter field theory of the problem. In one space dimension, this theory can be solved
analytically, giving the asymptotically exact critical behavior. Our paper is organized on the following
grounds: In Sec.~\ref{sec:m}, we introduce the order-parameter field theory and discuss the phase diagram.
Sec.~\ref{sec:SDRG} is dedicated to the derivation of the strong-disorder renormalization group flow equations
which are then analyzed and solved in Sec.~\ref{sec:flow_eqns}. In Sec.~\ref{sec:observables},
we compute the behavior of a few key observables. Finally, we conclude in Sec.~\ref{conc}.

\section{Order-parameter field theory}
\label{sec:m}

Following Landau, one of the general techniques for studying quantum phase transitions consists in deriving
an effective field theory in terms of the order parameter fluctuations from the the appropriate microscopic
Hamiltonian. For quantum phase transitions in Fermi liquids, this approach was pioneered by Hertz and
Millis \cite{Hertz76,Millis93}.
Our starting point is the resulting quantum Landau-Ginzburg-Wilson free energy functional (or action),
\begin{equation}
S=\int{\rm d}y{\rm d}x ~\varphi(x)\Gamma(x,y)\varphi(y)+\frac{u}{2N}\int{\rm d}x~\varphi^{4}(x).
\label{eq:clean-action}
\end{equation}
Here, $\varphi$ represents an $N$-component vector order-parameter in $d$ space dimensions. $x \equiv (\mathbf{x},\tau)$
stands for both the position vector $\mathbf{x}$ and imaginary time $\tau$ with
the integration measure $\int dx \equiv \int d^dx \int_0^{1/T} d\tau$. $u$ is the standard quartic
coefficient. In the absence of disorder, the two-point vertex $\Gamma(x,y)$ can be
defined via its Fourier transform which reads
\begin{equation}
\Gamma(\mathbf{q},\omega_{n})=r+\xi_{0}^{2}\mathbf{q}^{2}+\gamma_0\left|\omega_{n}\right|^{2/z_{0}},
\label{eq:2-pt}
\end{equation}
where $r$ represents the bare distance to criticality, and $\xi_0$ is a microscopic length scale.
The non-analytic dependence of $\Gamma(\mathbf{q},\omega_{n})$ on the Matsubara frequency $\omega_n$
is caused by the coupling of the order parameter field to a dissipative bath. We parameterize the dissipation
by the mean-field dynamical exponent $z_0$. If the dissipation is caused by simple oscillator baths
of spectral density $\mathcal{E}(\omega) \sim \omega^s$, one obtains $s=2/z_0$.
We are mostly interested in the case of superohmic
dissipation corresponding to $1< z_0 < 2$. However, for comparison, we also consider the dissipationless case,
$z_0=1$ and the Ohmic case, $z_0=2$. (Even weaker damping, formally corresponding to $z_0<1$, is irrelevant
because the resulting frequency term is subleading to the regular $\omega_n^2$ term arising in the Landau
expansion). The coefficient $\gamma_0$ derives from the the coupling of the order parameter to the dissipative
baths.
We now consider quenched disorder which does not change the functional form of the
Landau-Ginzburg-Wilson functional (\ref{eq:clean-action}) qualitatively. Instead,
the distance from criticality $r$ becomes a random function of the position vector
$\mathbf{x}$. Disorder also appears in the coefficients $\xi_0, \gamma_0$, and $u$.

Landau-Ginzburg-Wilson theories of the type defined in (\ref{eq:clean-action}) appear in several important physical
systems. For instance, the case of an $O(3)$ order parameter with Ohmic damping in $d=3$ dimensions
represents a quantum anti-ferromagnetic transition in itinerant electron systems \cite{Hertz76},
while an $O(2)$ order parameter
in one space dimension is relevant for the superconductor-metal transition in nanowires \cite{SachdevWernerTroyer04}.

In the following, we focus on the case of one space dimension and study the critical behavior of the model as
a function of the dissipation exponent $z_0$. We start by analyzing the phase diagram.  It is well known that in
sufficiently low dimensions, the effects of fluctuations are so large that they destroy any long-range order, as was rigorously
shown by Mermin and Wagner \cite{MerminWagner66}. To determine for which values of $z_0$ our model can display
long-range order, we calculate the strength of the transverse fluctuations in a putative ordered phase. Provided
that the dissipation remains active in the ordered phase\footnote{This is a nontrivial point. In some cases, the
action (\ref{eq:clean-action}) is only valid in the disordered phase, while a gap opens in the spectrum of the bath
modes in the ordered phase.}, the strength of the transverse fluctuations can be estimated by
\begin{equation}
\int_0^\Lambda d\omega\int dq \frac{1}{\xi_0^2 q^2+\gamma_0 \omega^{2/{z_0}}} = \int_0^\Lambda d\omega \frac {\pi}{\xi_0 \gamma_0^{1/2}  \omega^{1/z_0}}
\label{eq:mw1}
\end{equation}
where $\Lambda$ is a frequency cutoff. In the undamped case, $z_0=1$, the integral is logarithmically divergent in the infrared
limit. This implies that true long-range order is impossible but suggests a state with quasi-long-range order. In contrast,
for $z_0>1$, the integral is infrared convergent, implying that a long-range ordered phase can exist.

Having established the phase diagram, we now turn to the critical behavior of the order parameter
field theory (\ref{eq:clean-action}) in the presence of quenched disorder.
We aim at applying a strong-disorder renormalization group scheme. As this is a real-space based method, we
discretize the continuum action (\ref{eq:clean-action}) in space while imaginary time remains continuous.
Following Ref.\ \cite{VojtaKotabageHoyos09}, we define discrete spatial coordinates $\mathbf{x}_j$ and rotor
variables $\varphi_j(\tau)$ that represent the average order parameter in a volume 	$V$ large
compared to the microscopic scale $\xi_0$ but small compared to the true correlation length $\xi$.
In addition, we first consider the large-$N$ limit of the action (\ref{eq:clean-action}); we will later show that the
results remain valid for all continuous-symmetry cases.
The discretized version of the model in the large-$N$ limit and in $1+1$ dimensions reads as
\begin{eqnarray}
S & = & T\sum_{i}\sum_{\omega_{n}}\left(r_{i}+\lambda_{i}+\gamma_{i}\left|\omega_{n}\right|^{2/z_{0}}\right)\left|\phi_{i}(\omega_{n})\right|^{2}\nonumber \\
 &  & -T\sum_{ i }\sum_{\omega_{n}}\phi_{i}(-\omega_{n})J_{i}\phi_{i+1}(\omega_{n})\,.
 \label{eq:action}
 \end{eqnarray}
Here, $r_{i}$, $\gamma_{i}>0$ and the nearest-neighbor couplings $J_{ij}>0$ are random,
and $\phi_{j}(\omega_{n})=\int_{0}^{1/T}\varphi_{j}(\tau)e^{i\omega_{n}\tau}{\rm d}\tau$ is
the Fourier transform of the rotor variable. The Lagrange multipliers $\lambda_{i}$
enforce the large-$N$ constraints $\langle(\varphi_{i}^{(k)}(\tau))^{2}\rangle=1$ for
each of the order parameter components $\varphi_{i}^{(k)}$ at each site $i$; they have to be
determined self-consistently. The renormalized local distance from criticality at site
$i$ is given by $\epsilon_{i}=r_{i}+\lambda_{i}$. Note that in the disordered phase, \emph{all}
$\epsilon_i>0$.

\section{Strong Disorder Renormalization Group }
\label{sec:SDRG}
We now implement a strong-disorder renormalization group for the discrete large-$N$ action (\ref{eq:action}).
This method is based on successively integrating out local high-energy degrees of freedom
\cite{Fisher92,Fisher95,MaDasguptaHu79,DasguptaMa80,IgloiMonthus05}. The initial steps (until the derivation
of the recursion relations) closely follow Ref.\ \cite{VojtaKotabageHoyos09}, we will therefore only outline
the salient points of the derivation.

\subsection{Single-cluster solution}
\label{subsec:single_cluster}

A single independent rotor (or cluster) can be described by the action
\begin{equation}
S_{\rm cl} = T \sum_{\omega_n}\left(r+\lambda + \gamma|\omega_n|^{2/z_0} \right)
|\phi(\omega_n)|^2~. \label{eq:single-site-action}
\end{equation}
The Lagrange multiplier $\lambda$ is determined self-consistently by the length constraint
which at zero temperature reads
\begin{equation}
1= \langle \phi^2 \rangle  = \frac 1 {2\pi} \int_{-\infty}^{\infty} d\omega \frac 1 {\epsilon + \gamma
|\omega_n|^{2/z_0}}~, \label{eq:single-site-constraint-integral}
\end{equation}
where $\epsilon=r + \lambda$ is the renormalized distance from criticality. In
the super-Ohmic case, $z_0<2$, the integral can be carried out straight-forwardly,
and solving for the renormalized distance from criticality yields
\begin{equation}
\epsilon \sim \gamma^{z_0/(z_0-2)}~. \label{eq:gap-super-Ohmic}
\end{equation}
This power-law relation between $\epsilon$ and $\gamma$ (which is proportional to the cluster size)
shows that a single cluster is below the lower critical ``dimension'' of the problem.
In contrast, a high-frequency cutoff $\Lambda$ is needed to carry out the
constraint integral in the Ohmic case $z_0=2$. The resulting dependence of the gap on
the damping constant $\gamma$ is exponential,
$\epsilon =  \gamma \Lambda e^{-\pi\gamma}$,
signifying that a single Ohmic cluster is marginal, i.e., right at the lower critical
dimension of the problem.

\subsection{Recursion Relations}
\label{subsec:sd1}

In the action (\ref{eq:action}), the $\epsilon_i$ and $J_i$ are the competing local energy scales;
they are independent random variables with distributions $\pi(J)$ and $\rho(\epsilon)$.
In each renormalization step, we identify the largest of all of these energies, $\Omega=\max(\epsilon_i, J_i)$,
and then decimate the corresponding high-energy mode perturbatively. This procedure relies on the disorder
distributions being broad and becomes exact in the limit of infinitely broad distributions. For now, we assume
that our distributions are sufficiently broad; we verify this condition a posteriori.

Suppose the largest of all energies is a local distance from criticality (``gap''), say $\epsilon_2$, then we integrate out
the rotor $\varphi_2$ in a second-order perturbation expansion with the un-perturbed action given by $S_{0}=T\sum_{\omega_{n}}(\epsilon_{2}+\gamma_{2}\left|\omega_{n}\right|^{2/z_{0}})|\phi_{2}(\omega_{n})|^{2}$, and the
perturbation given by $S_{1}=-T\sum_{\omega_{n}} [J_{1}\phi_{1}(-\omega_{n})\phi_{2}(\omega_{n}) + J_{2}\phi_{2}(-\omega_{n})\phi_{3}(\omega_{n})]$.
This calculation is identical to that of the Ohmic case \cite{VojtaKotabageHoyos09} because the leading terms
do not depend on the value of $z_0$. It leads to the
term $\tilde{S}=-T\sum_{\omega_n}\tilde{J} \varphi_{1}(\omega_n) \varphi_{3}(-\omega_n)$ in the renormalized action
coupling sites 1 and 3.
The renormalized interaction strength $\tilde{J}$ is given by:
\begin{equation}
\tilde{J}= \frac{J_{1} J_{2}}{\epsilon_2}\,.
\label{eq:J-renorm}
\end{equation}
Note that this recursion relation is independent of the exponent $z_0$ which controls the strength of the dissipation.
Its multiplicative structure is fixed by second-order perturbation theory.

Suppose now that the dominant energy in the system is an interaction, say $J_{2}$ coupling the sites 2 and 3. In this limit,
the two rotors $\varphi_2$, and $\varphi_3$ are essentially locked together and can be replaced by a composite rotor
variable (``cluster'') $\tilde{\varphi_2}$. The resultant effective distance from criticality $\tilde{\epsilon_ 2}$ for this rotor can be
obtained by solving the two-site problem involving the rotor variable $\varphi_2$ and $\varphi_3$,
\begin{eqnarray}
\fl \quad
S_{0} &=& ~T\sum_{\omega_{n}}\sum_{i=2,3}(r_i+\lambda_i+\gamma_{i}\left|
\omega_{n}\right|^{2/z_{0}})|\phi_{i}(\omega_{n})|^{2} -
T\sum_{\omega_{n}}J_{2}\phi_{2}(-\omega_{n})\phi_{3}(\omega_{n})~.
\label{eq:two-site-action}
\end{eqnarray}
exactly whilst treating the coupling to the rest of the sites as perturbations. This calculation is straight forward but lengthy,
details will be shown in the appendix of this paper. In the superohmic case, $z_0<2$, the resulting recursion relation for $\tilde{\epsilon}_2$ reads:
\begin{equation}
{\left(\frac{\Omega}{\tilde{\epsilon}_2}\right)}^{x} = {\left(\frac{\Omega}{\epsilon_2}\right)}^{x} + {\left(\frac{\Omega}{\epsilon_3}\right)}^{x}  +{\rm const}
\label{eq:e-renorm}
\end{equation}
Here, the quantity $x=(2-z_0)/z_0$ vanishes in the Ohmic limit and takes the value 1 in the dissipationless case. The extra
constant approaches $-1$ in the Ohmic limit.
The new damping parameter is simply the sum of the two old ones,
\begin{equation}
\tilde\gamma_{2}=\gamma_{2}+\gamma_{3}~.
\label{eq:gamma-tilde}
\end{equation}
The additive structure of the recursion (\ref{eq:e-renorm}) in the superohmic case needs to be contrasted
with the multiplicative form $\tilde\epsilon = 2\epsilon_2 \epsilon_3/J_2$ arising in the Ohmic case
\cite{HoyosKotabageVojta07,VojtaKotabageHoyos09}. It is also worth mentioning that in the dissipationless
case, $z_0=1$, the recursion (\ref{eq:e-renorm}) agrees with that found in the dirty boson problem \cite{AKPR04},
as expected.

\subsection{Renormalization group flow equations}
\label{subsec:sd2}

The renormalization group steps outlined above are now repeated many times, thereby gradually reducing the maximum energy
$\Omega$. As a result, the probability distributions $\pi(J)$ and $\rho(\epsilon)$ change with $\Omega$.
The renormalization group flow equations for these distributions can be easily derived following Fisher's
treatment of the random transverse-field Ising model \cite{Fisher95}. This leads to
\begin{eqnarray}
\fl
-\frac{\partial \pi(J)}{\partial \Omega} &=& \left(\pi_{\Omega} - \rho_{\Omega}\right)\pi(J) + \rho_{\Omega}\int
dJ_1\;dJ_2\; \pi(J_1)\:\pi(J_2)\,  \delta\left(J-\frac{J_1J_2}{\Omega}\right),
\label{P-flow} \\
\fl
-\frac{\partial \rho(\epsilon)}{\partial \Omega} &=& \left(\rho_{\Omega} - \pi_{\Omega}\right)\rho(\epsilon)
+ \pi_{\Omega}\int
d\epsilon_1\;d\epsilon_2\; \rho(\epsilon_1)\:\rho(\epsilon_2)\, \delta\left(\epsilon^{-x}-\epsilon_1^{-x} - \epsilon_2^{-x} - \textrm{const}\right).
\label{R-flow}
\end{eqnarray}
Here, $\pi_\Omega$ and $\rho_\Omega$ denote the values of the distributions at their upper cutoff $\Omega$.
The flow equation for $\pi(J)$ is identical to the Ohmic case. However, the flow equation
for $\rho(\epsilon)$ is modified to reflect the additive recursion relation (\ref{eq:e-renorm}).

We now perform a set of variable transformations for $\Omega, J$ and $\epsilon$
\begin{equation}
\Gamma = \ln\frac{\Omega_0}{\Omega}, \quad \zeta = \ln\frac{\Omega}{J}, \quad \beta = \frac{1}{x}\left[
{\left(\frac{\Omega}{\epsilon}\right)}^x-1\right]
\label{eq:transformations}
\end{equation}
such that the recursion relations depicted (\ref{eq:J-renorm}) and (\ref{eq:e-renorm}) become simple additions.
Note that the transformation for $\beta$ turns into a logarithm in the Ohmic limit $x\to 0$.
Recast in the new variables, the flow equations for the distributions $P(\zeta)$ and $R(\beta)$ read
\begin{eqnarray}
\fl
-\frac{\partial P(\zeta)}{\partial \Gamma} &=& \frac{\partial P}{\partial \zeta} + \left(P_{0} - R_{0}\right)P
+ R_{0}\int
d\zeta_1\;d\zeta_2\; P(\zeta_1)\:P(\zeta_2) \delta\left(\zeta-\zeta_1-\zeta_2\right),
\label{P-mod-flow}
\\
\fl -\frac{\partial R(\beta)}{\partial \Gamma} &=& \left(1+ x\beta\right)\frac{\partial R}{\partial \beta} - \left(P_{0} - R_{0}-x\right)R
\nonumber\\
\fl &+& P_{0}\int
d\beta_1\;d\beta_2\; R(\beta_1)\:R(\beta_2) \delta\left(\beta-\beta_1-\beta_2-g(x) \right)
\label{R-mod-flow}
\end{eqnarray}
where $g(x)$ is a constant of order one. $P_0$ and $R_0$ denote the values of the
distributions at zero argument. In the Ohmic limit, the flow equations (\ref{P-mod-flow}) and (\ref{R-mod-flow})
reduce to those derived in Refs.\ \cite{HoyosKotabageVojta07,VojtaKotabageHoyos09} while in the dissipationless
case ($x=1$) they become identical to those of Altman et al. \cite{AKPR04} with $g(1)=1$.

\section{Solutions of the flow equations}
\label{sec:flow_eqns}

\begin{figure}
\begin{center}
\includegraphics[width=7.5cm,clip]{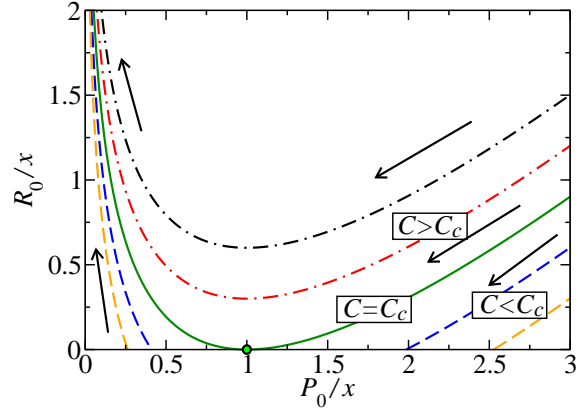}
\end{center}
\caption{Schematic of the renormalization group flow in the $P_0$-$R_0$ plane. For $C<C_c$, the
flow asymptotically approaches a line of stable fixed points that represent the ordered phase.
In the disordered phase, $C>C_c$, the flow is towards $P_0 \rightarrow 0$, and $R_0 \rightarrow \infty$.
For $C=C_c$, one flows into the critical fixed point at $P_0^*=x$, and $R_0^*=0$.}
\label{Fig:fd1}
\end{figure}

The complete set of fixed points of the flow equations (\ref{P-mod-flow}) and (\ref{R-mod-flow}) can be found utilizing
the methods developed in Refs.\ \cite{Fisher95,Fisher94}. We have performed such an analysis, details
will be published elsewhere \cite{MNHV10}. Here, we instead use the following ansatz \cite{AKPR04} for the
functional forms of the distributions $P(\zeta)$, and $R(\beta)$:
\begin{eqnarray}
P(\zeta) &=& P_0(\Gamma)\, e^{-P_0(\Gamma) \, \zeta}, \quad
R(\beta) = R_0(\Gamma)\,  e^{-R_0(\Gamma)\, \beta}
\label{eq:a1}
\end{eqnarray}
This exponential ansatz fulfils the flow equations (\ref{P-mod-flow}) and (\ref{R-mod-flow}) for the  distributions
$P(\zeta)$ and $R(\beta)$. The problem is thus reduced to flow equations for the parameters $P_0(\Gamma)$ and $R_0(\Gamma)$
which read
\begin{eqnarray}
\frac{dP_0}{d\Gamma} &=& -R_0 P_0~,
\label{eq:KT1}
\\
\frac{d R_0}{d \Gamma} &=& R_0 \left( x - P_0\right)~.
\label{eq:KT2}
\end{eqnarray}
It is interesting to note that these equations take the famous Kosterlitz-Thouless form \cite{KosterlitzThouless73}
for all $x>0$, i.e., for the entire range of superohmic dissipation strengths. (In fact, by a simple rescaling, the
flow for any $x>0$ can be mapped onto the $x=1$ case.) A schematic picture of the resulting
renormalization group flow in the $P_0$-$R_0$ plane in shown in Fig.\ \ref{Fig:fd1}.

There is a line of fixed points with $R_0^*=0$ and $P_0^*$ arbitrary. These fixed points are stable
for $P_0^*>x$ and unstable for $P_0^* < x$. The renormalization group trajectories in the $P_0$-$R_0$ plane
can be found by combining (\ref{eq:KT1}) and (\ref{eq:KT2}). This yields
\begin{equation}
R_0 = P_0 - x\ln P_0+C.
\label{eq:KT3}
\end{equation}
The integration constant $C$ controls in what phase the system is. For values of $C$ that are larger than a critical
value $C_c = -x +\ln x$ the flow is always towards $P_0 \rightarrow 0$, and
$R_0 \rightarrow \infty$. This implies that we asymptotically flow into a regime with $J \ll \epsilon$, i.e.,
the disordered phase. In the opposite regime, i.e., when $C<C_c$, the flow is towards a line of stable fixed points with
$P_0 =$ finite, and $R_0=0$. This corresponds to the ordered phase with  $\epsilon \ll J$.
The critical flow, corresponding to  $C=C_c$, separates the two regimes. It terminates at the
critical fixed point located at $P_0^*= x$, and $R_0^*= 0$.

We now linearize (\ref{eq:KT1}) and (\ref{eq:KT2}) around the critical fixed point,
to study the renormalization group flow in more detail. The critical flow ($C=C_c$) then takes the form
\begin{eqnarray}
R_0(\Gamma) &=& \frac{2}{x\Gamma^2}~, \quad P_0(\Gamma) = x+ \frac{2}{\Gamma}~.
\label{eq:KT4}
\end{eqnarray}
The density of surviving clusters at scale $\Gamma$ can be found by integrating the loss due to decimations
along the renormalization group flow. This yields
\begin{equation}
N(\Gamma) = N_0 e^{-\int_{\Gamma_0}^{\Gamma} d\Gamma^{\prime} \left[P_0(\Gamma^{\prime})+R_0(\Gamma^{\prime})\right]}
\label{eq:cl-1}
\end{equation}
By substituting the solutions (\ref{eq:KT4}) into this equation, one finds that the number of surviving clusters decreases
as $N(\Omega) \sim \Omega^x$ with decreasing cutoff energy $\Omega$. The typical distance $\ell$ between these clusters thus increases
as $\ell(\Omega) \sim \Omega^{-x}$. This implies that the dynamical scaling at the critical point is of power-law type.
The dynamical exponent $z=1/x$ interpolates continuously between the dissipationless case, $z=1$, and the Ohmic case
whose activated dynamical scaling formally corresponds to $z=\infty$.

It is important to realize that close to the critical fixed point, the interactions $J$ are much larger
than the gaps $\epsilon$. Therefore, almost all renormalization group steps will merge adjacent clusters,
while no clusters are eliminated. The number of sites (original rotors) contained in the surviving clusters
therefore remains essentially constant with $\Gamma \to \infty$. As a result, the typical size (moment) $\mu(\Omega)$ of
a surviving cluster scales inversely with their density, $\mu(\Omega)\sim \Omega^{-x}$.

The correlation length $\xi$ close to the critical point can be determined by considering the linearized flow equations off criticality. $\xi$ corresponds
to the length scale at which the off-critical flow significantly deviates from the critical one. In this way, we find the famous
exponential Kosterlitz-Thouless form
\begin{equation}
\ln(\xi/\xi_0) \sim {\sqrt{x/|r|}}
\label{eq:KT-xi}
\end{equation}
where $r = C-C_c$ denotes the (renormalized) distance from criticality.

Finally, we turn to the disordered Griffiths phase ($C>C_c$). Because $P_0$ flows to zero and $R_0$ to infinity
in this regime, we need to use the full flow equations (\ref{eq:KT1}) and (\ref{eq:KT2}) rather than the
linearized ones to determine the low-energy renormalization group flow. Solving (\ref{eq:KT1}) and (\ref{eq:KT2})
in the appropriate limit gives
\begin{eqnarray}
P_0(\Gamma) = P_x \exp\left[- (R_x/x)\, e^{x\Gamma}\right]~, \quad R_0(\Gamma) = R_x\, e^{x\Gamma}
\label{eq:Griffiths_flow}
\end{eqnarray}
where $P_x = \exp(C/x)$ and $R_x$ is an integration constant.
In the disordered Griffiths phase $J \ll \epsilon$, the system therefore consists of essentially independent
clusters with a distribution of characteristic energies $\epsilon$. Their density of states can be obtained by
inserting (\ref{eq:Griffiths_flow}) into (\ref{eq:a1}) and transforming back to the original distribution $\rho(\epsilon)$.
This yields
\begin{equation}
\rho(\epsilon) \sim \exp \left[ -(R_x/x) (\Omega_0/\epsilon)^x \right]
\label{eq:Griffiths_DOS}
\end{equation}
with subleading power-law corrections. We conclude that the Griffiths singularities in the superohmic case
are exponentially weak, in contrast to the Ohmic case which leads to strong power-law Griffiths singularities.
It is worth noting that the functional form of the rare region density of states (\ref{eq:Griffiths_DOS}) within our renormalization group
approach agrees with that predicted by heuristic optimal fluctuation arguments \cite{VojtaSchmalian05,VojtaSchmalian05b}.

\section{Observables}
\label{sec:observables}

To find the average order parameter, we run the renormalization group to $\Gamma \to \infty$ and
determine the total moment $\mu$ of the surviving cluster\footnote{We emphasize that the strong-disorder
renormalization group approach does not contain spin-wave type excitations of the ordered phase.
Whether or not an infinite surviving cluster indeed supports long-range order depends on the
strength of the corresponding transverse fluctuations, as discussed in Sec.\ \ref{sec:m}.}.
In the disordered phase, $C>C_c$, the flow is towards $P_0=0$ and $R_0=\infty$. Thus, all clusters
are decimated in the late stages of the renormalization group flow. The total moment of the surviving
clusters is therefore zero, implying that the average order parameter $m$ vanishes, as expected.
At criticality, the moment of a surviving cluster scales as $\mu(\Omega) \sim \Omega^{-x}$ while
their density behaves as $n(\Omega) \sim \Omega^x$ (see Sec.\ \ref{sec:flow_eqns}). The average
order parameter is given by $m = n(\Omega)\, \mu(\Omega)$ in the limit $\Omega \to 0$. We conclude
that $m$ is nonzero at the critical point.
To discuss how the order parameter changes upon moving into the ordered phase, we note that the
dependence of $m$ on $r=C-C_c$ stems from the rare renormalization group steps that decimate
clusters in the approach to the fixed point. By comparing the resulting losses of clusters along the
critical and off-critical renormalization group flows, we again find Kosterlitz-Thouless type behavior,
$m(r) - m(0) \sim \sqrt{|r|/x}$. In summary, the order parameter jumps from zero to a nonzero value
upon reaching the critical point and then increases in a square-root fashion.

The dependence of the order parameter susceptibility on temperature $T$ can be found by running the renormalization group
to $\Omega=T$ and treating the surviving clusters as independent because their interactions $J$ are small
compared to $T$. (Note that this constitutes an approximation in the superohmic case because the distribution
$\pi(J)$ does not become infinitely broad at criticality, but it becomes asymptotically exact for $x\to 0$.)
The total susceptibility is a sum of independent Curie contributions from the surviving clusters,
$\chi(T) = n(T) \mu^2(T)/T \sim T^{-(1+x)}$.

To calculate the spin-wave stiffness $\rho_s$, we apply boundary conditions at the two ends of the chain,
$x=0$ and $x=L$, such that the rotors at the two ends are at a relative angle $\Theta$. In the limit of
small $\Theta$ and large $L$, the free energy density $f$ depends on $\Theta$ as
$f(\Theta) -f(0) = \frac 1 2   \rho_s
\left( {\Theta}/{L}\right)^2$
which defines $\rho_s$. By optimizing the local twists according to the local interaction strengths,
one finds \cite{MohanNarayananVojta10} $\rho_s \sim \langle 1/J \rangle^{-1}$ where
$\langle \cdots \rangle$ denotes the average over $\pi(J)$. A full calculation would need to include
the internal stiffnesses of the surviving clusters. However, we can obtain an upper bound for the stiffness
by treating the surviving clusters as rigid. On the line of fixed points $P_0^* \ge x, R_0^*=0$
characterizing the critical point and the ordered phase, the distribution $\pi(J)$ follows from
transforming (\ref{eq:a1}) back to the original variables. It reads $\pi(J) = P_0^*\Omega^{-P_0^*} J^{P_0^*-1}$.
The integral $\int_0^\Omega dJ \, \pi(J)/J$ diverges at the lower bound for all $P_0 \le 1$.
We thus conclude that the spin-wave stiffness $\rho_s$ vanishes not just at the critical point ($P_0^*=x$),
but also in part of the ordered phase (for $x<P_0^*\le 1$).

\section{Conclusions and Discussions}
\label{conc}
In summary, we have studied the influence of superohmic dissipation on the quantum critical behavior of a disordered $O(N)$ $\phi^4$
order parameter field theory. We have shown that the static critical behavior falls in the Kosterlitz-Thouless universality class independent
of the exponent that controls the dissipation strength. The dynamical scaling is of conventional power-law type with the
dynamical exponent $z$ varying from 1 in the dissipationless case to infinity in the limit of Ohmic dissipation.
In both limiting cases, our theory agrees with known results. For $z_0=1$
we reproduce the dirty boson results of Ref.\ \cite{AKPR04}, and in the limit $z_0=2$ our theory becomes identical
to the Ohmic case \cite{HoyosKotabageVojta07,VojtaKotabageHoyos09}.
In these final paragraphs, we discuss a number of remaining questions,
and we put our results into a broader perspective.

\begin{figure}
\begin{center}
\includegraphics[width=7.5cm,clip]{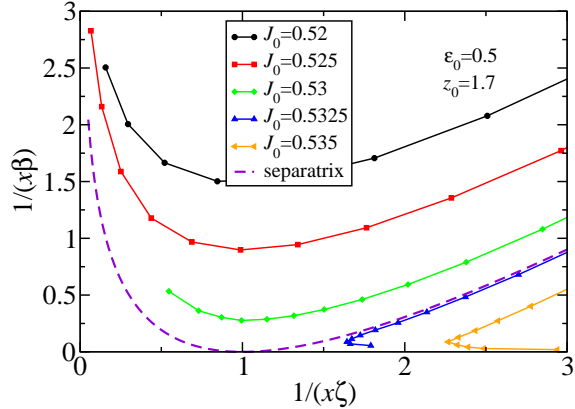}
\end{center}
\caption{Numerical implementation of the strong-disorder renormalization group: Trajectories of the average
         values of $\zeta$ and $\beta$ along the renormalization group flow. The bare interactions $J$ are distributed
         uniformly between $J_0$ and 1; the bare gaps are distributed uniformly between $\epsilon_0$ and 1.
         The exponent $z_0=1.7$. The dashed line is the critical trajectory predicted by (\ref{eq:KT3}).
         The right-turns of the lowest two curves for very small $1/(x\beta)$ are a numerical artefact.
 }
\label{Fig:flow_num}
\end{figure}

First, while the exponential ansatz (\ref{eq:a1}) provides a solution of the renormalization group flow equations, it is a priori
not clear that the flow for a moderately disordered system actually follows this solution. We have therefore implemented the
recursion relations (\ref{eq:J-renorm}) and (\ref{eq:e-renorm}) numerically and studied the renormalization group flow
for a broad range of parameters.  A characteristic result is shown in Fig.\ \ref{Fig:flow_num}, more details will be published
elsewhere \cite{MNHV10}.
The figure shows that the flow agrees very well with that predicted by our analytical solution.
(Note that within the exponential ansatz (\ref{eq:a1}), the average values of $\zeta$ and $\beta$
are given by $P_0^{-1}$ and $R_0^{-1}$, respectively.)

Second, the recursion relations (\ref{eq:J-renorm}) and (\ref{eq:e-renorm}) are valid for broad disorder
distributions. Close to the critical fixed point, the distribution $\rho(\epsilon)$ does become infinitely
broad ($R_0 \to 0$). The recursion relation (\ref{eq:e-renorm}) for merging two clusters thus becomes
asymptotically exact.  Because essentially all renormalization steps close to the critical fixed point
consist of merging two clusters, as discussed in Sec.\ \ref{sec:flow_eqns}, we conclude that the critical
fixed point is stable against subleading corrections to the recursion relations. To test the effects of these
corrections during the early stages of the renormalization group, we have included them in the numerical
solution. We found that they do not qualitatively change the flow towards the fixed points discussed in
the present paper \cite{MNHV10}.

We also point out that our explicit calculations are for the large-$N$ limit of the order-parameter field
theory (\ref{eq:clean-action}). However, the critical behavior depends only on the structure of the recursion
relations (\ref{eq:J-renorm}) and (\ref{eq:e-renorm}). The multiplicative structure of (\ref{eq:J-renorm})
simply reflects second-order perturbation theory and will certainly be the same for finite $N$.
The structure of (\ref{eq:e-renorm}) reflects the nontrivial power-law dependence of $\epsilon$ on the
cluster size for superohmic dissipation. It is the same for all continuous-symmetry cases, as also follows
from a renormalization group analysis of appropriate nonlinear sigma model at its strong-coupling
fixed point \cite{VojtaSchmalian05b}. We thus conclude that the critical behavior found here will be
valid for all $N>1$.

The theory developed in this paper applies to one space dimension. In higher dimensions, a closed
analytical solution of the strong-disorder renormalization group appears impossible because the topology
of the lattice changes under the renormalization procedure. However, a few qualitative aspects of
the behavior in higher dimensions can be read off the structure of the theory. In particular, the fact
that we obtained power-law dynamical scaling rather than activated scaling arises from the
power-law relation (\ref{eq:gap-super-Ohmic}) between the ``gap'' (distance from criticality)
and the damping parameter of a single cluster. The damping parameter is proportional to the number of sites in
the cluster and thus, in general, to some power of the cluster length scale, $\gamma \sim L^{D_f}$
where $D_f$ is the (fractal) dimension of a surviving cluster at criticality. Together with
(\ref{eq:gap-super-Ohmic}), this suggests that the dynamical exponent is generally given by
$z= D_f \, z_0/(2-z_0)$. Interestingly, this is precisely the relation found at a percolation
quantum phase transition \cite{VojtaSchmalian05b} where $D_f$ is simply the geometric
fractal dimension of the underlying percolating lattice. The power-law relation (\ref{eq:gap-super-Ohmic})
between the gap and the damping parameter also implies that the Griffiths singularities are exponentially
weak in all dimensions because, when combined with the exponential size distribution of rare regions,
it produces an exponentially small low-energy density of states \cite{VojtaSchmalian05}.
Even though the dynamical critical behavior in higher dimensions will be similar to our one-dimensional
results, we do not expect the Kosterlitz-Thouless character to survive in higher dimensions. This follows
from the observation that in $d>1$, even the dissipationless quantum rotor system has a conventional
quantum phase transition.

Finally, we relate our results to the general classification of phase transitions in the presence of disorder
based on the defect (and rare region) dimensionality \cite{Vojta06,VojtaSchmalian05}. As can be seen from the
calculations in Sec.\ \ref{subsec:single_cluster}, a single cluster
or rare region is \emph{below} the lower critical dimension of the problem
in the case of superohmic dissipation. This puts the transition into class A of the classification,
which implies conventional power-law dynamical scaling and exponentially weak Griffiths singularities.
The results of Sec.\ \ref{sec:flow_eqns} are in complete agreement with these predictions.

\ack

We acknowledge helpful discussions with G. Refael. This work has been supported in part by the NSF under grant
nos.\  DMR-0339147, PHY-0551164, and DMR-0906566, by Research Corporation, by FAPESP under grant No. 2010/03749-4,
by CNPq under grant No. 302301/2009-7, and by DST grant number PHY/10-11/232/DSTX/ALAK.
Parts of this work have been performed at the Kavli Institute for Theoretical
Physics, Santa Barbara and at the Aspen Center for Physics.

\appendix
\section*{Appendix}
\setcounter{section}{1}

In this appendix, we sketch the derivation of the recursion relations  (\ref{eq:e-renorm}) and
(\ref{eq:gamma-tilde}) for the case of a strong bond. To identify and integrate out the
high-energy mode of the two-site cluster, we diagonalize the quadratic form in its action (\ref{eq:two-site-action}).
This yields
\begin{equation}
S_0 = T\sum_{\omega_n} \left[ \kappa_a |\psi_a(\omega_n)|^2 + \kappa_b |\psi_b(\omega_n)|^2 \right]~.
\label{eq:S_0_diagonalized}
\end{equation}
Here, $\psi_a$ and $\psi_b$ are the two eigenmodes,
and the corresponding eigenvalues are given by
\begin{eqnarray}
\kappa_{a,b} = \frac 1 2 \left( d_2+d_3 \mp \sqrt{(d_2-d_3)^2 +J_{2}^2} \right)
\label{eq:eigenvalues}
\end{eqnarray}
with $d_j =r_j +\lambda_j +\gamma_j |\omega_n|^{2/z_0}$. Expanding in $1/J_2$ results in
\begin{eqnarray}
\fl
\kappa_{a,b} = \frac 1 2 \left[ r_2+\lambda_2 +r_3 +\lambda_3 \mp J_2 +(\gamma_2+\gamma_3)|\omega_n|^{2/z_0} \right] + O\left(\frac 1 {J_2}\right)~.
\label{eq:eigenvalues_expanded}
\end{eqnarray}
As the higher eigenvalue $\kappa_b$ is at least $J_2$ above the lower eigenvalue
$\kappa_a$, we integrate out the eigenmode $\psi_b$ yielding the effective
action $\tilde{S}_0 = T\sum_{\omega_n} \kappa_a |\psi_a(\omega_n)|^2$. We now rescale
the remaining eigenmode by defining $\tilde{\phi}_2 = \psi_a / \langle \psi_a^2 \rangle^{1/2}$
because we wish the renormalized rotor to fulfill the same large-$N$ constraint
$\langle \tilde{\phi}_2^2 \rangle=1$ as the original variables. In the relevant limit
$J_2 \gg \epsilon_2, \epsilon_3$, the rescaling factor $\langle \psi_a^2 \rangle$
approaches 2. We thus arrive at the renormalized action
\begin{equation}
\fl
\tilde{S}_0 = T\sum_{\omega_n} \left[ r_2 +\lambda_2 +r_3 +\lambda_3 - J_2 +(\gamma_2 +\gamma_3) |\omega_n|^{2/z_0} \right] |\tilde{\phi}_2(\omega_n)|^2~.
\label{eq:S_0_renormalized}
\end{equation}
From the frequency term, we can immediately read off the recursion relation
(\ref{eq:gamma-tilde}), $\tilde{\gamma}_2=\gamma_2 + \gamma_3$, for the damping
coefficients. If we combine this recursion with the single cluster relation
(\ref{eq:gap-super-Ohmic}) between $\gamma$ and $\epsilon$, we obtain
the recursion relation (\ref{eq:e-renorm}) for the local distance from criticality.

Alternatively (and to check the consistency of the procedure), one can calculate
the renormalized distance from criticality directly from the two-site cluster solution
via $\tilde{\epsilon}_2 = r_2 +\lambda_2 +r_3 +\lambda_3 - J_2$. This requires finding
the Lagrange multipliers $\lambda_2$ and $\lambda_3$ from the
large-$N$ constraint equations
\begin{eqnarray}
1&=& \langle \phi_2^2 \rangle  = T\sum_{\omega_n}  \frac {d_3}{d_2 d_3-J^2/4}~,
    \nonumber \\
1&=& \langle \phi_3^2 \rangle  = T\sum_{\omega_n} \frac {d_2}{d_2 d_3-J^2/4}~.
\label{eq:twosite-constraint}
\end{eqnarray}
We have performed this calculation in complete analogy to the Ohmic case
\cite{VojtaKotabageHoyos09} and rederived the recursion relation (\ref{eq:e-renorm}) for
the renormalized distance from criticality $\tilde{\epsilon}_2$. The result is identical to that
obtained via the renormalized damping constant $\tilde{\gamma}_2$ above.

\section*{References}
\bibliographystyle{iopart-num}
\bibliography{../00Bibtex/rareregions}

\providecommand{\newblock}{}
\begin{thebibliography}{10}
\expandafter\ifx\csname url\endcsname\relax
  \def\url#1{{\tt #1}}\fi
\expandafter\ifx\csname urlprefix\endcsname\relax\def\urlprefix{URL }\fi
\providecommand{\eprint}[2][]{\url{#2}}

\bibitem{ThillHuse95}
Thill M and Huse D~A 1995 {\em Physica A\/} {\bf 214} 321

\bibitem{RiegerYoung96}
Rieger H and Young A~P 1996 {\em Phys. Rev. B\/} {\bf 54} 3328

\bibitem{Fisher92}
Fisher D~S 1992 {\em Phys. Rev. Lett.\/} {\bf 69} 534

\bibitem{Fisher95}
Fisher D~S 1995 {\em Phys. Rev. B\/} {\bf 51} 6411

\bibitem{Vojta03a}
Vojta T 2003 {\em Phys. Rev. Lett.\/} {\bf 90} 107202

\bibitem{Vojta06}
Vojta T 2006 {\em J. Phys. A\/} {\bf 39} R143

\bibitem{Vojta10}
Vojta T 2010 {\em J. Low Temp. Phys.\/} {\bf 161} 299

\bibitem{CastroNetoJones00}
Castro~Neto A~H and Jones B~A 2000 {\em Phys. Rev. B\/} {\bf 62} 14975

\bibitem{MillisMorrSchmalian01}
Millis A~J, Morr D~K and Schmalian J 2001 {\em Phys. Rev. Lett.\/} {\bf 87}
  167202

\bibitem{MillisMorrSchmalian02}
Millis A~J, Morr D~K and Schmalian J 2002 {\em Phys. Rev. B\/} {\bf 66} 174433

\bibitem{SchehrRieger06}
Schehr G and Rieger H 2006 {\em Phys. Rev. Lett.\/} {\bf 96} 227201

\bibitem{SchehrRieger08}
Schehr G and Rieger H 2008 {\em J. Stat. Mech.\/}  P04012

\bibitem{HoyosVojta08}
Hoyos J~A and Vojta T 2008 {\em Phys. Rev. Lett.\/} {\bf 100} 240601

\bibitem{HoyosKotabageVojta07}
Hoyos J~A, Kotabage C and Vojta T 2007 {\em Phys. Rev. Lett.\/} {\bf 99} 230601

\bibitem{VojtaKotabageHoyos09}
Vojta T, Kotabage C and Hoyos J~A 2009 {\em Phys. Rev. B\/} {\bf 79} 024401

\bibitem{Hertz76}
Hertz J 1976 {\em Phys. Rev. B\/} {\bf 14} 1165

\bibitem{SachdevWernerTroyer04}
Sachdev S, Werner P and Troyer M 2004 {\em Phys. Rev. Lett.\/} {\bf 92} 237003

\bibitem{VojtaSchmalian05}
Vojta T and Schmalian J 2005 {\em Phys. Rev. B\/} {\bf 72} 045438

\bibitem{AKPR04}
Altman E, Kafri Y, Polkovnikov A and Refael G 2004 {\em Phys. Rev. Lett.\/}
  {\bf 93} 150402

\bibitem{Millis93}
Millis A~J 1993 {\em Phys. Rev. B\/} {\bf 48} 7183

\bibitem{MerminWagner66}
Mermin N~D and Wagner H 1966 {\em Phys. Rev. Lett.\/} {\bf 17} 1133

\bibitem{MaDasguptaHu79}
Ma S~K, Dasgupta C and Hu C~K 1979 {\em Phys. Rev. Lett.\/} {\bf 43} 1434

\bibitem{DasguptaMa80}
Dasgupta C and Ma S~K 1980 {\em Phys. Rev. B\/} {\bf 22} 1305

\bibitem{IgloiMonthus05}
Igloi F and Monthus C 2005 {\em Phys. Rep.\/} {\bf 412} 277

\bibitem{Fisher94}
Fisher D~S 1994 {\em Phys. Rev. B\/} {\bf 50} 3799

\bibitem{MNHV10}
Mohan P, Narayanan R, Hoyos J and Vojta T Unpublished

\bibitem{KosterlitzThouless73}
Kosterlitz J~M and Thouless D~J 1973 {\em J. Phys. C\/} {\bf 6} 1181

\bibitem{VojtaSchmalian05b}
Vojta T and Schmalian J 2005 {\em Phys. Rev. Lett.\/} {\bf 95} 237206

\bibitem{MohanNarayananVojta10}
Mohan P, Narayanan R and Vojta T 2010 {\em Phys. Rev. B\/} {\bf 81} 144407

\end{thebibliography}

\end{document}